\documentclass[a4paper]{mn2e}

\usepackage{epsfig}

\title[Red giant depletion in globular cluster cores]{Red giant
depletion in globular cluster cores}

\author[M. E. Beer and M. B. Davies]{Martin
E. Beer\thanks{E-mail: martin.beer@astro.le.ac.uk} and Melvyn
B. Davies\\Department of Physics and Astronomy, University of Leicester,
Leicester, LE1 7RH, United Kingdom}

\date{23rd January 2003}

\volume{000}

\pagerange{\pageref{firstpage}--\pageref{lastpage}} \pubyear{2003}

\begin{document}

\maketitle

\label{firstpage}

\begin{abstract}
We investigate the observed depletion of red giants in the cores of
post-core-collapse globular clusters. In particular, the evolutionary
scenario we consider is a binary consisting of two low-mass stars
which undergoes two common envelope phases. The first common
envelope phase occurs when the primary is a red giant resulting in a
helium white dwarf and main sequence star in a detached binary. The second
common envelope phase occurs shortly after the 
secondary becomes a red giant. During the second common envelope phase
the degenerate helium cores merge resulting in a core mass greater
than the helium burning limit and the formation of a horizontal branch
star. We show that this evolutionary route is enhanced in
post-core-collapse clusters by stellar encounters. These encounters
increase the population of binary secondaries which would have evolved
onto the red giant branch in the recent past.
\end{abstract}

\begin{keywords}
binaries: close -- stars: evolution -- stars: general -- globular
clusters: general.
\end{keywords}

\section{Introduction}
Observations by Djorgovski et al. (1991) have shown that clusters with
central cusps (presumably collapsed cores) have colour and population 
gradients. Their observations show that a number of post-core-collapse
(PCC) clusters become bluer toward their centres, which is an effect
of the demise of the red giant and/or sub-giant populations, and
possibly an increase in the number of faint, blue objects. The effect
is a few per cent of the visible light.

The cores of PCC clusters have been imaged using the
high spatial resolution of the Hubble Space Telescope (Bailyn 1994;
Shara et al. 1998). The observations show a depletion of the red giant
branch stars relative to the horizontal branch.
In addition a population of supra-horizontal-branch stars (SHBs) have
been observed. These are stars which lie in the
Hertzsprung-Russell diagram as a distinct population brighter than
normal horizontal branch stars. Bailyn (1994) found SHBs in the core
of 47 Tuc which were about one magnitude brighter, while Shara et
al. (1998), in images of NGC 6522, found SHBs which were one and a half
magnitudes brighter in the $B$ band. The formation of these objects is
not well understood.

In PCC clusters more stellar encounters are expected to occur than
for the non-PCC clusters. These encounters could deplete the red giant
population through two methods. Firstly, collisions between red giants
and binaries could either eject the red giant core from the surrounding
envelope or result in a common envelope phase (Adams, Sills \& Davies
2003). This effect could at most deplete a few per cent of the red
giant population. Secondly stellar encounters would modify the binary
population through hardening (encounters making the binary tighter)
and exchanges (the third star is exchanged with the lower mass binary
component if it is more massive). This would result in a different
binary population to the non-PCC clusters. This modified population
evolves such that the red giants, which would be observed today,
undergo mergers with their binary companions.

As red giants evolve into horizontal branch stars any evolutionary
model which is considered needs to explain why the horizontal branch
is not depleted in the same manner as the red giants. We
investigate whether such an evolutionary scenario is possible. Such a
scenario would require binary interactions to turn/accelerate the
evolution of the red giants into becoming horizontal branch stars (see
Djorgovski et al. 1991). In this paper we discuss an evolutionary scenario
which naturally explains both the observed depletion of red giants and
the non-depletion of horizontal branch/faint, blue stars in PCC
clusters. In this scenario, the binary consists of two 
low-mass (below helium-flash limit) stars and evolves through
two common envelope phases. In the first common envelope phase the
primary is a red giant and loses its envelope to leave a helium white
dwarf--main sequence star binary. During the second common envelope phase
, in which the secondary has evolved into a red giant, the two
degenerate helium cores coalesce resulting in a helium burning core
and the remainder of the common envelope forming an envelope around
the merged object. 

We examine the required parameter space (initial masses and
separation) to explain why the depletion is more likely to occur in
PCC clusters. Stellar encounters are expected to play a significant
role in PCC clusters. After a number of encounters the original binary
will have had its separation ground down and its components will have 
higher masses due to exchanges (see Davies 1995). The initial mass
range which would produce red giants today is between 0.75 M$_{\sun}$
and 0.95 M$_{\sun}$ depending on the age of the cluster and hence the
turn-off mass. We show that without encounters this mass range is more
likely to be found in single stars and the higher mass binary
components (the primaries). After a number of encounters more are
found in binaries and in particular as the secondaries of binaries. It
is these secondaries which lose their envelopes in the second common
envelope phase explaining why red giant depletion is only observed in
PCC clusters.

Our evolutionary scenario is discussed in section~\ref{evolscen}. In
section~\ref{encsect} we investigate the effects stellar encounters
have in modifying the binary population of PCC
clusters. Section~\ref{discuss} contains a discussion of our results,
in particular the observational consequences and the other products of
our evolutionary scenario. We give our conclusions in
section~\ref{conclude}.

\section{Evolutionary scenario} \label{evolscen}
\begin{figure}
\begin{center}
\epsfig{file=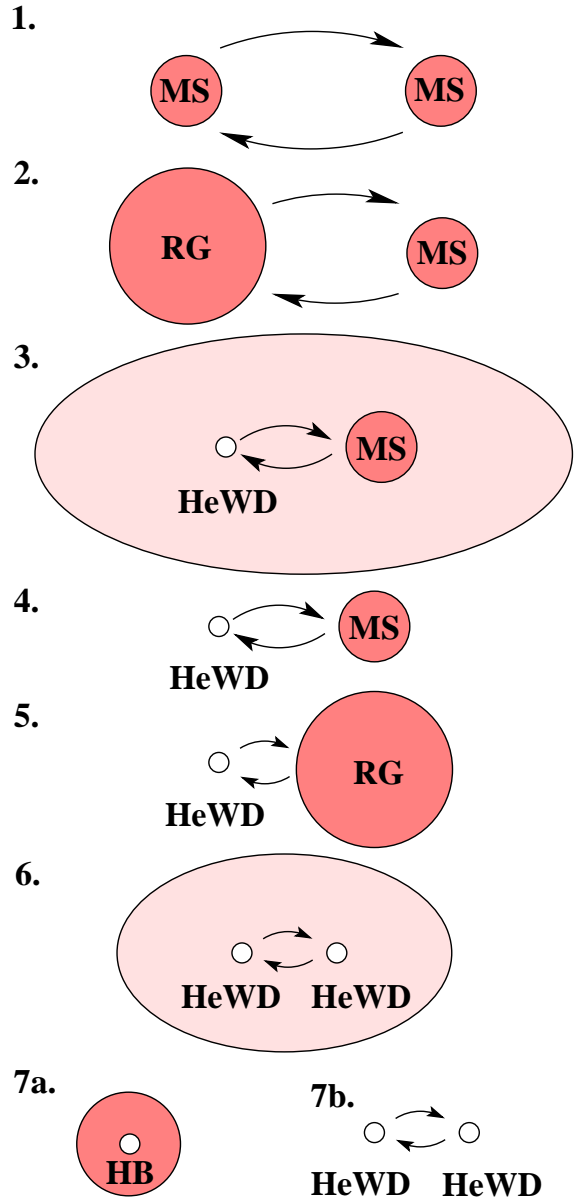,width=7.5cm}
\caption{An evolutionary scenario for depleting the red giant branch,
some of which may become horizontal branch stars. Beginning with two
main-sequence stars (phase 1), the primary evolves into a red giant
(phase 2) filling its Roche lobe, leading to the onset of a common
envelope phase (3) where the red-giant envelope engulfs the
main-sequence secondary star and the degenerate helium core. The
post-common-envelope system consists of a tight binary containing the
degenerate helium core (a white dwarf) and the main-sequence secondary
star (phase 4). The main-sequence star evolves into a red giant (phase 5)
filling its Roche lobe, leading to the onset of a second common
envelope phase (6) where the red-giant envelope engulfs the helium
white dwarf and the degenerate helium core. If coalescence occurs in
the common envelope, then a degenerate helium core is formed
surrounded by the remainder of the common envelope. If it is
sufficiently massive to ignite helium ($M_{\rm c}>0.48~{\rm
M}_{\sun}$) it may appear as a horizontal branch star (phase 7a)
otherwise it will appear as a red giant. Alternatively, after the
second common envelope phase a tight double degenerate helium white
dwarf system would be formed (phase 7b).} \label{figevol}
\end{center}
\end{figure}

The evolutionary scenario for depleting the red giant branch is shown
in Fig.~\ref{figevol}. We consider the evolution of binary stars where
convective mass transfer occurs when the primary evolves off the main
sequence and fills its Roche lobe. A common envelope phase
follows. The secondary and the degenerate helium core spiral together
as the envelope is ejected. The secondary then evolves to fill its
Roche lobe and convective mass transfer again occurs forming a second
common envelope phase. During this common envelope phase the
degenerate helium cores spiral together. 

If the cores coalesce during the second common envelope phase then the
combined core masses may be enough to ignite helium and the remainder of
the common envelope is left around the merged helium burning core and
resembles a horizontal branch star. If the hydrogen envelope mass is
small then the product may resemble an sdB star. Alternatively, after
the second common envelope phase a tight double degenerate helium
white dwarf system could be formed. This system could be close enough
that inspiral due to gravitational radiation brings them into
contact.

To model the binary evolution we use the binary evolution code of
Hurley, Tout \& Pols (2002). This code uses analytical formulae for
stellar evolution (Hurley, Pols \& Tout 2000) and includes the effects
of tides, gravitational radiation, mass transfer and mass loss due to
winds. For the common envelope the code assumes the inspiral is given
by an energy balance between the envelope binding energy and the
difference in orbital energy of the inspiraling components. The
envelope binding energy is given by
\begin{equation}
E_{\rm bind} = -\frac{G M_{\rm c 1} M_{\rm e n v 1}} {\lambda R_1} ~,\
\end{equation}
where $M_{\rm c 1}$ and $M_{\rm e n v 1}$ are the core and envelope mass
of the red giant respectively, $R_1$ is its radius and $\lambda$ is a
constant. The change in orbital binding energy is given by
\begin{equation}
\delta E_{\rm orb} = -\frac{1}{2} \left ( \frac{G M_{\rm c 1} M_2}
{a_{\rm f}} - \frac{G M_{\rm c 1} M_2} {a_{\rm i}} \right ) ~,\
\end{equation}
where $a_{\rm i}$ and $a_{\rm f}$ are the initial and final
separations respectively and $M_2$ is the mass of the secondary. The
energy balance is then given by
\begin{equation}
E_{\rm bind} = \alpha_{\rm C E} ~ \delta E_{\rm orb} ~,\
\end{equation}
where $\alpha_{\rm C E}$ is the common envelope efficiency. The ratio
of initial to final separation only depends on the product
$\alpha_{\rm C E} \,\lambda$ and not on their separate values. We take
the value of $\alpha_{\rm C E} \,\lambda = 1$ in our calculations.

\subsection{Constraints on initial masses}
\begin{figure}
\begin{center}
\epsfig{file=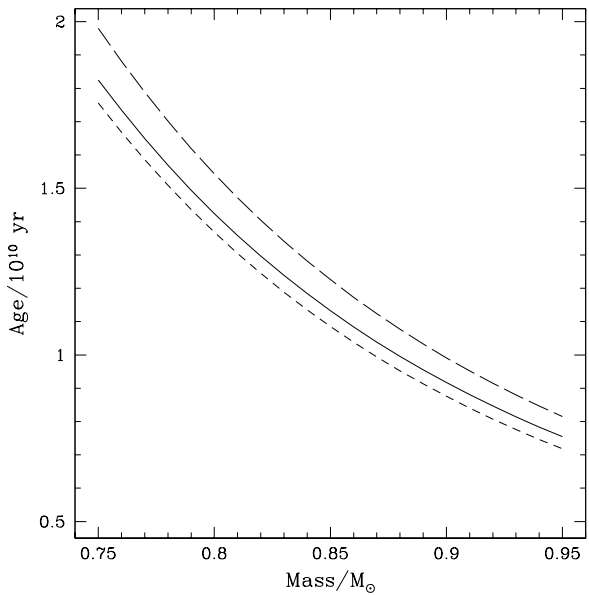,width=8.4cm}
\caption{A figure showing the time to reach the end of the red giant
branch as a function of the initial age for different
metallicities. The solid line, short-dashed line and long dashed line
are for metallicities of 0.001, 0.0001 and 0.002 respectively.}
\label{figmturn}
\end{center}
\end{figure}

In order for the observed red giant population to be depleted the
secondaries in the binaries require masses near the turn-off mass for
the cluster. Fig.~\ref{figmturn} shows the age of a star at the end
of the red giant branch  as function of initial mass and for
metallicities z = 0.0001, 0.001, 0.002.
From Fig.~\ref{figmturn} it is clear that a large range of cluster
ages of 10 -- 20 Gyr can be represented if we consider a mass range
for the secondary of 0.75 -- 0.95~M$_{\sun}$. This represents the
range of masses of secondaries which would be observed as red giants
in globular clusters today.

\subsection{Constraints on initial separations}
The initial separation needs to be such that both binary
components overflow their Roche lobes as red giants. If the separation
is too small then the secondary will overflow its Roche lobe before it
reaches the red giant branch. This will deplete both the red giant and
horizontal branch. Alternatively if the separation is too
large then the binary will not come into contact for a second time. In
this case no depletion of either the red giant or the horizontal
branch occurs.

\begin{figure}
\begin{center}
\epsfig{file=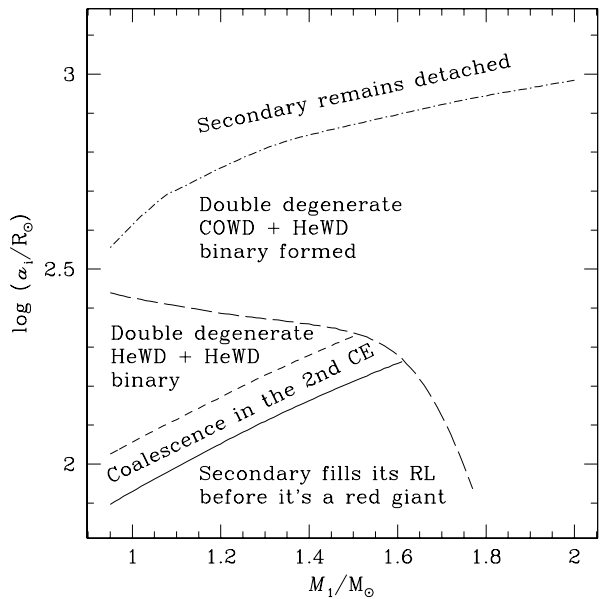,width=8.4cm}
\caption{Plot of initial separation $a_{\rm i}$, for $M_2$ = 
0.8~M$_{\sun}$, as a function of primary mass $M_1$ showing constraints
for the formation of a binary as described in section~\ref{evolscen}.}
\label{figm1var}
\end{center}
\end{figure}

Fig.~\ref{figm1var} is a plot of initial primary masses and
separations and shows the constraints for the formation of a
horizontal branch star via the coalescence of two degenerate helium
cores during the second common envelope phase. We assume a metallicity
of 0.001 and a secondary mass of 0.8 M$_{\sun}$. The solid line shows
the separation above which coalescence occurs in the second common
envelope. Below this separation the secondary fills its Roche lobe
before it has reached the red giant branch. The short-dashed line
represents the separation at which a double degenerate helium white
dwarf system is formed as opposed to coalescence occurring during the
common envelope. The long-dashed line represents the separation at
which a helium white dwarf and carbon-oxygen white dwarf is formed due
to the primary not filling its Roche lobe until the asymptotic giant
branch. The dot-dashed line represents the separation above which the
secondary does not fill its Roche lobe.

\begin{figure}
\begin{center}
\epsfig{file=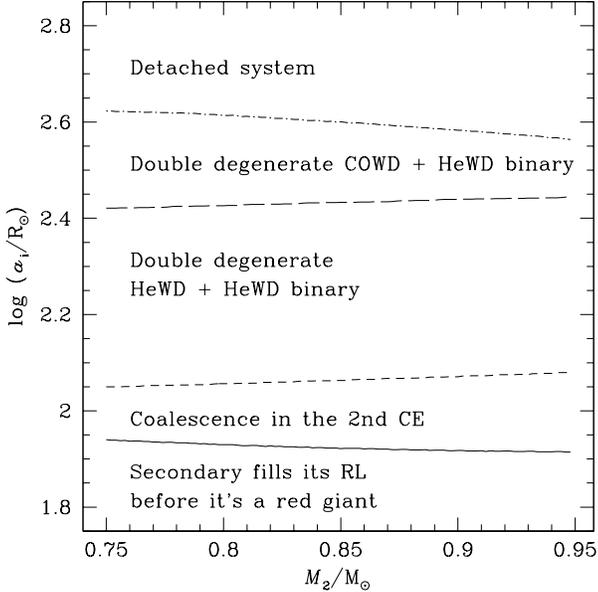,width=8.4cm}
\caption{Plot of initial separation $a_{\rm i}$, for $M_1$ = 
1~M$_{\sun}$, as a function of secondary mass $M_2$ showing constraints
for the formation of a binary as described in section~\ref{evolscen}.}
\label{figwhichwork}
\end{center}
\end{figure}

Fig.~\ref{figwhichwork} is a plot of initial secondary masses and
separations and shows the constraints for the formation of a
horizontal branch star via the coalescence of two degenerate helium
cores during the second common envelope phase. For the evolutionary
calculation we assumed a metallicity z = 0.001 and a primary mass of 1
M$_{\sun}$. The lines are the same as in
Fig.~\ref{figm1var}. Fig.~\ref{figwhichwork} shows that the
constraints for formation of the merged object are only weakly
dependent on the secondary mass.

In all of our calculations in which two degenerate helium cores merge
during the second common envelope we find that the combined core mass
is sufficient to ignite helium. In all the cases which form a
horizontal branch star, the primary overflows its Roche lobe near the
end of the giant branch. The core mass of the primary in this case is
around 0.4~M$_{\sun}$. For the second common envelope 
phase we require the secondary to be a red giant and consequently its
core mass is at least 0.2~M$_{\sun}$. Consequently the combined core
masses are always sufficiently large enough to ignite helium (which
requires $M_{\rm c}>0.48~{\rm M}_{\sun}$).

We have run the evolution code with different values for the
common envelope efficiency and neglecting the mass loss due to the
wind. Neglecting the wind did not significantly change the evolutionary
outcome of the systems and the range of parameter space was found to
be similar for each model. The common envelope efficiency, however,
determines the separations at which coalescence during the second
common envelope phase will occur. The dot-dashed line above which the
primary overflows its Roche lobe on the asymptotic giant branch is
unaffected but the separation after the first common envelope phase
is. This in turn affects whether a merger will occur in the second
common envelope phase as this will depend on both the common envelope
efficiency and the initial separation. It was found that decreasing
the efficiency by a factor two shifted the parameter space in which
coalescence occurs upwards by a factor 0.2 dex but without
significantly changing its area. The common envelope
efficiency consequently mainly affects the proportion of double
degenerates which are formed relative to the horizontal branch stars.
This demonstrates that the evolutionary scenario is a common outcome 
and not a consequence of the selection of certain parameters.

\section{Role of the stellar encounters in a post-core-collapse
cluster} \label{encsect}
\subsection{Binary hardening due to stellar encounters}
During stellar encounters the orbital separation is hardened as the
third body interacts with the system (Davies 1995). Initially
binaries will have a range of orbital separations but the wide
binaries will become tighter pushing more binaries into the parameter
space where they are likely to undergo two common envelope phases and
merge to form a horizontal branch star. As an example in the cluster
47 Tuc a 1000 R$_{\sun}$ binary can be ground down to 100 R$_{\sun}$
(Davies \& Benz 1995).

\subsection{Modification of binary masses due to stellar encounters}
\label{sectimf} 
\begin{figure}
\begin{center}
\epsfig{file=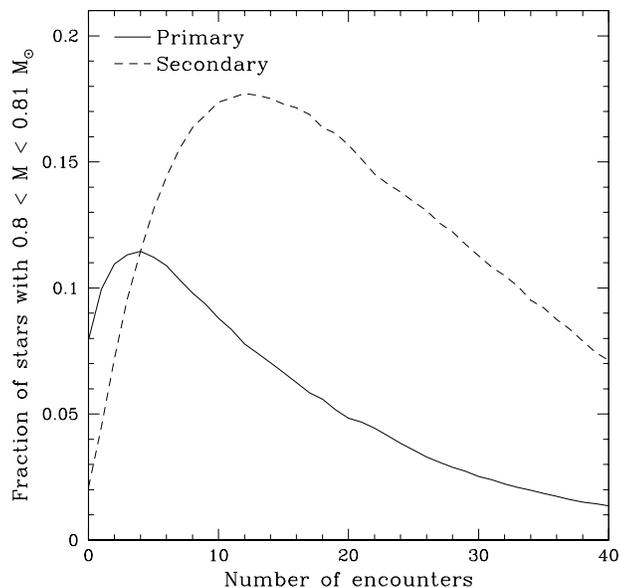,width=8.4cm}
\caption{A plot of the proportion of stars with masses between 0.8 and
0.81 M$_{\sun}$ in the different binary components as a function of
the number of stellar encounters a binary population has
experienced. The solid line is the primary population and the dashed
line is the secondary.} \label{figpop}
\end{center}
\end{figure}

When a star encounters a binary, the lowest mass star is usually
ejected (Davies 1995). This could be either the third star or
one of the binary components. The number of encounters depends on the
encounter lifetime ($\tau_{\rm enc}$) which is given by
\begin{equation}
\tau_{\rm enc} = 7 \times 10^4 \frac {n} {10^5 /{\rm p c}^3} \left (
\frac{{\rm M}_{\sun}} {M_1 + M_2} \right ) \frac {{\rm R}_{\sun}}
{b_{\rm min}} \frac {v_{\infty}} {10 \,{\rm km\,s^{-1}}} {\rm Myr}~,\
\end{equation}
where $n$ is the number density of stars, $b_{\rm min}$ is the
distance of closest approach and $v_{\infty}$ is the stellar
velocity. For typical parameters in the core of a PCC cluster,
$\tau_{\rm enc}$ is approximately 400 Myr. This implies that during
the main sequence lifetime of a 0.8 M$_{\sun}$ star we can expect
20--40 encounters. To determine how encounters change the population
of binary components we calculated the effect of a number of stellar
encounters on a binary population. We assumed an initial population of
stars with an initial mass function given by Eggleton, Fitchett \&
Tout (1989)
\begin{equation}
\frac{M}{M_{\sun}} = \frac{0.19x} {(1-x)^{0.75} + 0.032(1-x)^{0.25}}
\,
\end{equation}
where $x$ is a random number between 0 and 1. This appears as a
power-law for masses greater than 1~M$_{\sun}$ and with a turnover at
low masses (with a peak at $M\sim0.18~$M$_{\sun}$). Of our initial
population of stars, we assumed 10\% of them were in binaries. 
To mimic encounters we assumed that a binary has encounters with a
star randomly drawn from the single star population. During an
encounter if the third star is more massive than one of the binary
components they will be exchanged (see Davies 1995) i.e., the
remaining binary always consists of the two most massive stars. We
simulated a number of encounters for each binary to build up an image
of how the population changes with the number of encounters. The mass
range which would have evolved onto the red giant branch in the recent
past is 0.8 -- 0.81 M$_{\sun}$. Fig.~\ref{figpop} shows the fraction of
stars with this mass range which are the primary and secondary
components of binary systems as a function of the number of
encounters the binary population has experienced. The results did not
significantly change if a different range of masses was chosen.
Table~\ref{tableenc} shows how the proportions of primary 
and secondary components after 0, 20 and 40 encounters per binary. In
addition the table shows the proportion of stars which may form red
giant or horizontal branch stars. We assume only the single stars are
likely to form red giants while horizontal branch stars will form from
both the single stars and the secondaries of binary systems (through
mergers). In the case of the horizontal branch
these numbers are an upper limit and depend on the proportion of the
binary population which have the necessary masses and separations to
undergo a merger during the second common envelope phase. 
From Table~\ref{tableenc} and Fig.~\ref{figpop} it is clear that in a
non-PCC cluster the population of stars would be the same as the field
with roughly 8\% of the stars in the mass range of interest as the
primaries of binaries and almost none as the secondaries. Consequently
the red giant population will not be significantly depleted through
binaries undergoing two common envelope phases in a non-PCC cluster. 
After 20 encounters, however, the situation is reversed. The secondary
population has increased up to 16\% while the primary component has
dropped to 5\%. Consequently more systems can form horizontal branch
stars than red giants in a cluster whose members have had a number of
stellar encounters. After 40 
encounters, the secondary population has decreased to 7\% and the
primary population to 1\%. However, the red giant branch will still be
depleted relative to the horizontal branch.
\begin{table*}
\begin{center}
\begin{tabular}{l c c c c c}
\hline
N$_{\rm enc}$ & \multicolumn{3}{c}{Proportion of stars which are} &
\multicolumn{2}{c}{Proportion which can form} \\
& Single & Primary & Secondary & Red Giants & Horizontal Branch \\
\hline
0 & 90.0 \% & 7.9 \% & 2.1 \% & 90.0 \% & 92.1 \% \\
20 & 79.5 \% & 4.8 \% & 15.7 \% & 79.5 \% & 95.2 \% \\
40 &  91.5 \% & 1.4 \% & 7.1 \% & 91.5 \% & 98.6 \% \\
\hline
\end{tabular}
\caption{A table showing the proportion of stars, in the mass range
0.8 -- 0.81 M$_{\sun}$ which are single or the components of a binary
after 0, 20 and 40 encounters per binary system. During the
lifetime of a 0.8 M$_{\sun}$ star 20--40 encounters per binary are
expected to occur in the core of a PCC cluster. Also shown is the
proportion of these stars which will form the red giant and horizontal
branch population.} \label{tableenc}
\end{center}
\end{table*}

This naturally explains why the depletion of red giants is only
observed in PCC clusters. In non-PCC clusters there is not a
significant number of binaries which have secondaries of the
cluster turn-off mass. After a number of encounters, however, the
situation is reversed and a number of turn-off mass stars are the
secondaries in binaries. These binaries may undergo two common
envelope phases with coalescence in the second common envelope. These
mergers form horizontal branch stars and deplete the red giant
population.

\section{Discussion} \label{discuss}
We have shown that if a binary system undergoes two common envelope
phases it is possible for a merger of the two cores to occur. We have
assumed that this results in a horizontal branch star. The masses of
the cores will add up to greater than the mass required for helium
ignition. If all the helium ignited simultaneously then the nuclear
energy generated would be greater than the binding energy of the
star. However, as in the case of degenerate helium ignition at the end
of the red giant branch, the burning timescale is large enough
($\sim$100~Myr) that this does not occur and the envelope remains
bound.

If the cores merge near the end of the common envelope phase then the
final star will have ejected a large portion of the hydrogen envelope
and so the merged object may resemble an sdB star. Han et al. (2002)
have investigated the formation channels for sdB stars and although
one of the routes considered is dynamical mass transfer followed by a
common envelope phase, two common envelope phases are not
discussed (although the merger of two white dwarfs is one of the
formation channels considered). Whether or not this is a main
evolution channel is unknown as it may depend strongly on the initial
separation.

The merged objects in our calculations will have a higher
core mass and a lower envelope mass than normal horizontal branch
stars. The higher core mass will result in a larger luminosity and so
the SHBs observed by Bailyn (1994) and Shara et al. (1998) (or a
subset of them) may be the merged objects which would form after
coalescence during the second common envelope phase.

We have shown that in the cores of PCC clusters the conditions are
such that depletion of the red giant population is possible. We have
not, however, performed detailed binary population synthesis of the
cluster as a whole. This is complicated due to the necessity of
including stellar encounters, stellar evolution and binary
interactions. The solution to this may be the use of N-body modelling
codes. However all we are showing here is that this evolutionary route
is both possible and likely in PCC cluster cores. In this paper we do
not attempt the determination of the depletion due to common envelope
mergers including the effects of encounters, exchanges and binary
evolution simultaneously.

In addition to the horizontal branch stars formed through mergers our
calculations predict the existence of a substantial double degenerate
population. 
The double degenerate population we find in our calculations have two
main components. The first is where both are helium white dwarfs. In
this case the primary will have overflowed its Roche lobe near the end
of the red giant branch while the secondary will typically overflow
near the start of the red giant branch. This is a consequence of the
inspiral during the common envelope phase resulting in the radii of
the two stars at Roche lobe overflow being significantly
different. The mass ratio in this case is typically between 1.5 and 2
and the final orbital period is a few hours. 

The second component to the double degenerate population comes from
carbon-oxygen and helium white dwarf binaries. These are formed when
the primary overflows its Roche lobe on the asymptotic giant branch
resulting in both the carbon-oxygen white dwarf-helium white dwarf
binary and a larger separation. The mass ratio in these cases are
between 1 and 1.5 with larger periods of a couple of days
upwards. Comparing to the observed population of Maxted, Marsh \&
Moran (2002) we find that these most closely resemble their population
in which the mass ratios are around one and the periods are from a few
hours to slightly greater than a day. However, there is a significant
difference in the ages determined from cooling models. In their models
the white dwarfs are within a few 100 Myr of each other while in our
models the more massive primary overflows its Roche lobe typically a
few Gyr before the secondary. A possible remedy to this scenario may
be binaries which are closer in mass to each other or which initially
were much more massive. In both these cases they
would become giants in similar time-scales allowing the white dwarfs to
be formed closer in age to each other. The other populations of
double degenerates, however, should still be present and it may be
that there are selection effects which determine which binaries are
observed or that the double degenerate population in the absence of a
number of encounters is different..

Han (1998) has modelled the formation of double degenerates including
two common envelope phases. This evolutionary route was found to
produce a significant portion of the total double degenerate
population. Nelemans et al. (2001) have also modelled the population
of double degenerates although they use different assumptions
concerning the inspiral during the first common envelope phase and the
mass loss due to a wind.

Of the double degenerates which do form a number of them are close
enough that they inspiral and mass transfer occurs. In the case of
helium-helium white dwarfs the mass ratio is such that the mass
transfer is always stable. In this case as mass is transferred the
binary becomes wider again. The final binary in this case has a very
low mass secondary and a $\sim$0.6~M$_{\sun}$ helium white dwarf
primary which could undergo ignition to form a carbon oxygen white
dwarf. 

\section{Conclusions} \label{conclude}
We have described an evolutionary scenario in which the red giants are
depleted in the cores of PCC clusters. The binary undergoes two common
envelope phases during the second of which coalescence of two
degenerate helium cores occurs. The combined mass of these cores is
sufficient for helium burning to commence resulting in the formation
of horizontal branch stars. The binary separations required for this
to occur are around 100 R$_{\sun}$ for a primary of mass 1
M$_{\sun}$ and a range of secondary masses.

We have shown that this evolutionary scenario will be favoured in the
cores of PCC due to the effects of stellar encounters. These
encounters both harden the binary towards the required orbital separation
and increase the population of secondaries which have the cluster
turn-off mass.

\section*{Acknowledgements}
We thank Jarrod Hurley for providing his binary evolution
code. Theoretical astrophysics at Leicester is supported by a PPARC
rolling grant. MEB gratefully acknowledges a UK Astrophysical Fluids
Facility (UKAFF) fellowship.

\end{document}